\documentclass[aps,prl,nofootinbib,reprint,nobibnotes,showpacs,superscriptaddress,]{revtex4-1} 

\usepackage[english]{babel}
\addto\captionsenglish{

}

\usepackage{graphicx}
\graphicspath{{graphics/}}

\usepackage{subfigure}
\usepackage[shortlabels]{enumitem}
\usepackage{multirow}
\usepackage{float}

\usepackage{amsmath}
\usepackage{amssymb}
\usepackage[version=3]{mhchem}
\usepackage{mathrsfs}
\usepackage{bm}

\usepackage[hidelinks,linktocpage]{hyperref}
\hypersetup{
  colorlinks,
  linkcolor={red!70!black},
  citecolor={green!70!black},
  urlcolor={blue!70!black}
}

\usepackage[usenames,dvipsnames]{xcolor}
\usepackage{comment} 
\usepackage{slashed} 
\usepackage{cancel}
\usepackage[normalem]{ulem}

\newcommand{\nbb}{\ensuremath{0\nu\beta\beta} } 
\newcommand{\vvbb}{\ensuremath{2\nu\beta\beta} } 

\begin{document}

\title{New direct limit on neutrinoless double beta decay half-life of $^{128}$Te with CUORE}

\author{D.~Q.~Adams}
\affiliation{Department of Physics and Astronomy, University of South Carolina, Columbia, SC 29208, USA}

\author{C.~Alduino}
\affiliation{Department of Physics and Astronomy, University of South Carolina, Columbia, SC 29208, USA}

\author{K.~Alfonso}
\affiliation{Department of Physics and Astronomy, University of California, Los Angeles, CA 90095, USA}

\author{F.~T.~Avignone~III}
\affiliation{Department of Physics and Astronomy, University of South Carolina, Columbia, SC 29208, USA}

\author{O.~Azzolini}
\affiliation{INFN -- Laboratori Nazionali di Legnaro, Legnaro (Padova) I-35020, Italy}

\author{G.~Bari}
\affiliation{INFN -- Sezione di Bologna, Bologna I-40127, Italy}

\author{F.~Bellini}
\affiliation{Dipartimento di Fisica, Sapienza Universit\`{a} di Roma, Roma I-00185, Italy}
\affiliation{INFN -- Sezione di Roma, Roma I-00185, Italy}

\author{G.~Benato}
\affiliation{INFN -- Laboratori Nazionali del Gran Sasso, Assergi (L'Aquila) I-67100, Italy}

\author{M.~Beretta}
\affiliation{Department of Physics, University of California, Berkeley, CA 94720, USA}

\author{M.~Biassoni}
\affiliation{INFN -- Sezione di Milano Bicocca, Milano I-20126, Italy}

% ...
\date{\today}

\date{\today}

\begin{abstract}
  The Cryogenic Underground Observatory for Rare Events (CUORE) at Laboratori Nazionali del Gran Sasso of INFN in Italy
  is an experiment searching for neutrinoless double beta (0$\nu\beta\beta$) decay.
  Its main goal is to investigate this decay in $^{130}$Te,
  but its ton-scale mass and low background make CUORE sensitive to other rare processes as well.
  In this work, we present our first results on the search for \nbb decay of $^{128}$Te,
  the Te isotope with the second highest natural isotopic abundance.
  We find no evidence for this decay, and using a Bayesian analysis we set a lower limit
  on the $^{128}$Te \nbb decay half-life of T$_{1/2} > 3.6 \times 10^{24}$ yr (90\% CI).
  This represents the most stringent limit on the half-life of this isotope,
  improving by over a factor 30 the previous direct search results,
  and exceeding those from geochemical experiments for the first time.

\end{abstract}

\maketitle

\label{sec:intro}

Double beta ($\beta\beta$) decay is a rare second-order Fermi interaction
in which a nucleus $(A,Z)$ transforms into its isobar $(A,Z + 2)$
by the simultaneous transmutation of two neutrons into two protons.
This Standard Model process occurs with the emission of two electrons
and two electron antineutrinos in the final state (\vvbb decay),
such that lepton number ($L$) conservation holds;
this process has been measured for 11 nuclei~\cite{Barabash:2020nck},
with half-lives in the range of 10$^{18}$-10$^{22}$ years.
A second decay mode, neutrinoless double beta (0$\nu\beta\beta$) decay,
has been hypothesized but never observed.
This process would consist of a nucleus $\beta\beta$-decaying into its daughter
with the emission of two electrons and no antineutrinos in the final state,
thus violating $L$ by two units.
The experimental signature of this process is a peak in the two-electron total energy spectrum at the Q-value (Q$_{\beta\beta}$)
of the transition.
The search for \nbb decay addresses one of the most relevant open questions in neutrino physics:
its observation would establish that $L$ is not a symmetry of nature and neutrinos are Majorana fermions,
providing a clear signature of physics beyond the Standard Model~\cite{Bilenky:2014uka,Dolinski:2019nrj}.
This would provide significant input for the explanation of 
the matter-antimatter asymmetry in the Universe via leptogenesis~\cite{Canetti:2012zc,Luty:1992un},
as well as constraints 
on the absolute mass scale and ordering of neutrinos complementing other approaches~\cite{Dolinski:2019nrj,DellOro:2019pqi}.

CUORE is a ton-scale array of 988 TeO$_2$ crystals designed to search for \nbb decay of $^{130}$Te. Besides having the world leading sensitivity for this process~\cite{CUORE:2019yfd,CUORE:2021mvw} due to its very large mass --742\,kg of TeO$_2$-- and low background, CUORE is also a powerful detector for other rare processes, in particular other Te decay channels~\cite{CUORE:2020bok,CUORE:2021xns,Adams:2022lkq}. In this letter we report on a new direct search for $^{128}$Te {\nbb} decay. The CUORE array is grown from material with natural isotopic composition, which given the natural abundance of 31.75\%~\cite{Fehr200483} contains 188\,kg of $^{128}$Te.
Despite this high abundance, the direct search is challenging due to the low Q$_{\beta\beta}$ value of ($866.7\pm0.7)$\,keV~\cite{Wang:2021xhn} which lies in a region of the energy spectrum dominated by \vvbb decay of $^{130}$Te and $\gamma$ backgrounds from other natural radioactivity.  
The most recent $^{128}$Te $0\nu\beta\beta$ decay half-life limit from a direct search experiment, T$^{0\nu}_{1/2} > 1.1\cdot10^{23}$\,yr, was set by MiDBD in 2003~\cite{Arnaboldi:2002te}. More stringent limits than this have been set by indirect geochemical measurements (see~\cite{Campani:2021cfh} for a review), which evaluate the presence of the $\beta\beta$ decay products accumulated in geological mineral samples of known age via the assessment of the parent/daughter nuclei ratio. 
The geochemical studies are not sensitive to the $\beta\beta$ decay mode
but rather to the sum of all the possible decays (2$\nu\beta\beta$ or 0$\nu\beta\beta$, to the ground or excited states), although the dominant contribution is expected to be the two-neutrino mode. The direct search result reported in this letter improves by more than 30-fold the previous best direct search limit for this isotope and surpasses -- for the first time -- the indirect geochemical results. 

Before reporting the details of our direct search for $^{128}$Te \nbb decay, we present an updated evaluation of the half-life value for $^{128}$Te $\beta\beta$ decay based the ratio T$_{1/2}$($^{130}$Te)/T$_{1/2}$($^{128}$Te) = $(3.52\pm 0.11)\cdot 10^{-4}$~\cite{Bernatowicz:1992ma} from ion-counting mass spectrometry of Xe in ancient Te samples. Using the most recent $^{130}$Te \vvbb decay half-life measurement, $7.71^{+0.08}_{-0.06}$(stat.)$^{+0.12}_{-0.15}$(syst.)$\times \, 10^{20}$ yr~\cite{CUORE:2020bok},  we obtain
T$^{2\nu}_{1/2}(^{128}$Te$)=(2.19 \pm 0.07)\cdot 10^{24}$\,yr.
This result replaces and is in agreement with the previously published value of
$\mathrm{T^{2\nu}_{1/2}(^{128}Te)} = (2.25 \pm 0.09)\cdot 10^{24}$ yr~\cite{Barabash:2020nck},
which used the weighted average of the $^{130}$Te \vvbb decay half-lives from CUORE-0~\cite{CUORE:2016ons} and CUORE~\cite{Nutini:2020vtd}.

\label{sec:cuore}

The CUORE detector comprises 19 towers of 52 crystals each. The basic unit is a 5$\times$5$\times$5\,cm$^3$ TeO$_2$ crystal operated as an individual cryogenic calorimeter. Each crystal is equipped with a Neutron Transmutation Doped (NTD) Ge thermistor~\cite{Haller1984}, used as a temperature sensor, and a Si resistor to inject controlled heat pulses for thermal gain stabilization. The crystal is coupled through PTFE and Cu supports to the coldest stage of a dilution refrigerator operating at a temperature of $\sim$10 mK~\cite{Alduino:2019xia}.
Any particle interaction in a TeO$_2$ absorber crystal produces an energy deposition that is converted into heat (phonons) and measured via the temperature sensor.
A large and novel cryogenic infrastructure has been developed to provide the needed cooling power~\cite{CUORE:2021mvw}. The CUORE cryostat  
is designed to meet the CUORE background specifications~\cite{Alduino:2017qet}, and provide a low thermal noise environment, minimizing vibration and thermal dissipation on the cryogenic calorimeters~\cite{DAddabbo:2017efe,Dompe:2020,ADAMS2022103902,Alduino:2019xia}.
CUORE is the most advanced realization of the cryogenic calorimetric technology,
developed over 30 years using TeO$_2$-based detectors~\cite{Brofferio:2018lys}.

We acquire data in day-long periods called runs, which in turn are grouped into $\sim$40~--~60 day collections called datasets.
A typical dataset consists of 4~--~5 days of calibration runs,
followed by 30~--~50 days of so-called physics runs, and finally another 4~--~5 days
of final calibration to check the energy scale stability within a dataset.
Calibration runs are performed using $\gamma$-ray sources of $^{232}$Th and $^{60}$Co to illuminate the detectors.

The procedure for the data acquisition and processing is described in~\cite{CUORE:2019yfd}.
We apply a digital optimum trigger algorithm~\cite{DiDomizio:2010ph,Campani:2020ltd}
to the acquired continuous data stream and evaluate the amplitude of the triggered waveform
by applying a frequency-based Optimum Filter (OF) that weights the Fourier components of the signal,
exploiting the noise power spectrum to reduce the impact of noisy frequencies.
We compensate for thermal gain variations in the crystals due to small fluctuations
in their operating temperature with two independent methods.
The first utilizes heat pulses of fixed amplitude injected regularly (every 570 s) via the Si heaters affixed to the crystals.
For crystals with non-functional heaters we use the 2615\,keV $\gamma$ events from $^{208}$Tl in calibration data as a reference.
We use the data from calibration runs to convert the thermal amplitudes to units of energy.
We exploit the granularity of the CUORE detector to perform a coincidence study
and determine if signals in different crystals within a short time and spatial distance
(typically of 10 ms and 150 mm) are attributed to the same physical interaction.
We refer to these as coincident signals, to which we assign a multiplicity number,
$\cal{M}$$_n$, where $n$ corresponds to the number of crystals simultaneously involved in the interaction
(e.g., two events in different crystals due to Compton scattering of the 2615\,keV $^{208}$Tl line
are labeled as ${\cal{M}}_2$), with single-crystal interactions labeled as ${\cal{M}}_1$.
We apply a pulse shape analysis (PSA) algorithm to identify and discriminate pulses
due to particle energy depositions from non-physical signals (e.g., noise spikes, abrupt baseline disturbances, pile up events).

The present analysis includes 5 datasets for a total TeO$_2$ exposure of 309.33\,kg$\cdot$yr or 78.56\,kg$\cdot$yr of $^{128}$Te.
These are the same data we used to measure 
the $^{130}$Te \vvbb decay half-life~\cite{CUORE:2020bok}.
However, the latter exposure is marginally lower (300.72\,kg$\cdot$yr) due to stricter selection criteria
on the energy scale calibration in both the $\beta/\gamma$($<$3~MeV) and $\alpha$($>$3 MeV) regions for the \vvbb decay result.
In contrast, this analysis requires only good performance in the $\beta/\gamma$ region.

In the following, we provide a detailed description of the analysis technique used to search for $^{128}$Te \nbb\ decay,
whose signature is a mono-energetic peak at  Q$_{\beta\beta}=(866.7\pm0.7)$\,keV in the summed energy of the two emitted electrons.
In the great majority of the cases the two electrons are absorbed by the same crystal:
we therefore select ${\cal{M}}_1$ events only, within a region of interest (ROI) of (820~--~890)\,keV.

The signal efficiency is the product of the containment efficiency and the total analysis efficiency.
We define the containment efficiency ($\epsilon_{\text{MC}}$) as the fraction of $^{128}$Te \nbb decay events
that release their full energy, i.e. Q$_{\beta\beta}$, in a single crystal~\cite{CUORE:2016acf}.
We evaluate $\epsilon_{\text{MC}}$ by simulating $10^8$ events in the CUORE crystals~\cite{Alduino:2017qet},
obtaining $\epsilon_{\text{MC}}=97.59\pm0.01$\%.
The total analysis efficiency (exposure-weighted average $\epsilon_{cut} = 87.74\pm 0.19\%$) is the product of the total reconstruction efficiency,
the anti-coincidence efficiency and the PSA efficiency.
The first term is the probability that an event with a given energy is triggered,
its energy is correctly reconstructed, and it is not rejected as a pile-up event
by the analysis cuts applied during the data processing;
the anti-coincidence efficiency is the probability that a single-hit event
is not assigned the wrong multiplicity due to a random accidental coincidence with an unrelated event;
the PSA efficiency is the probability that events passing the base pile-up cuts also survive the PSA cut.
We refer to~\cite{CUORE:2019yfd} for a more detailed description of the computation methods of these efficiency terms.

\begin{figure}[h]
  \centering
  \includegraphics[width=1.\columnwidth]{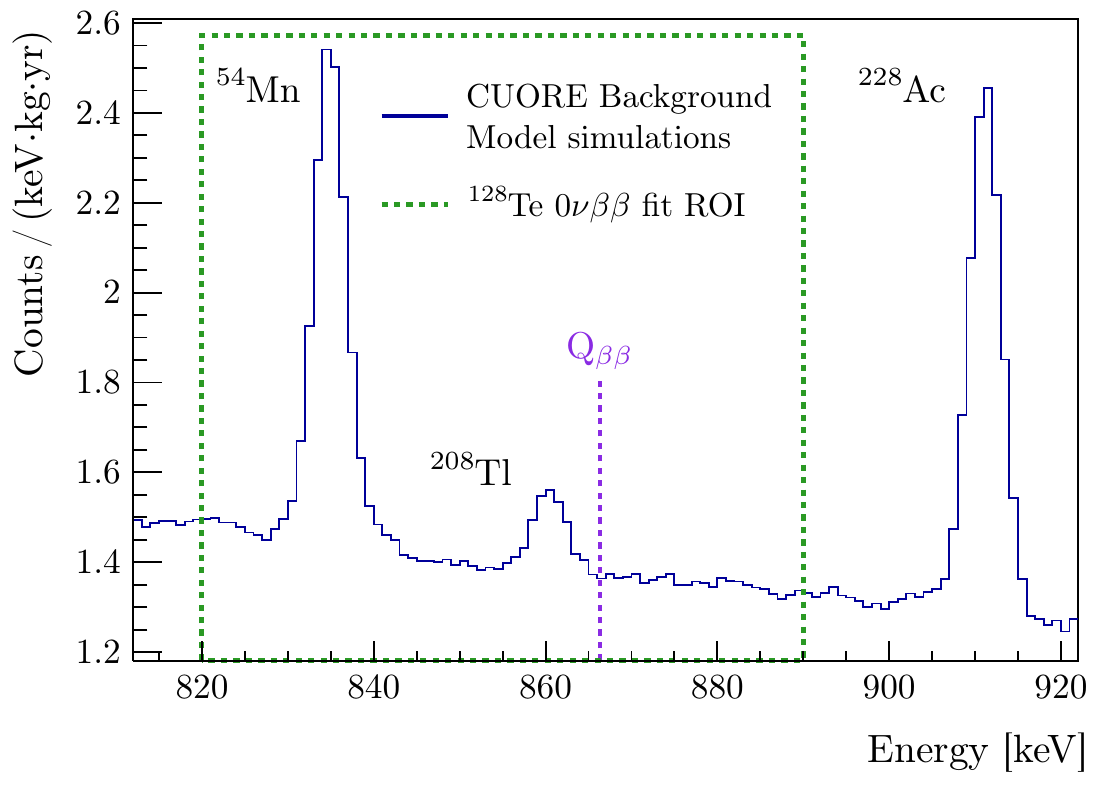}
  \caption{${\cal{M}}_1$ spectrum from the CUORE Background Model simulations
    in the proximity of the $^{128}$Te \nbb decay Q$_{\beta\beta}$.
    From left to right: $^{54}$Mn $\gamma$  (834.8\,keV), $^{208}$Tl $\gamma$  (860.6\,keV) and $^{228}$Ac $\gamma$  (911.2\,keV).
    The ROI for this analysis is denoted by the dashed green box, and includes the $^{54}$Mn and $^{208}$Tl lines.}
  \label{fig:ROI_BM}
\end{figure}

To avoid introducing bias when choosing the fit model of the present analysis,
we choose the ROI based on the CUORE Background Model (BM) simulations,
particularly taking into account backgrounds close to Q$_{\beta\beta}$ for $^{128}$Te (Fig. \ref{fig:ROI_BM}). 
Based on this, we choose an ROI of (820~--~890)\,keV.
 \\
Multiple peaks populate this energy window: the closest expected structure to Q$_{\beta\beta}$
is a $\gamma$ line at 860.6\,keV from $^{208}$Tl, a $^{232}$Th chain element.
A prominent peak at 834.8\,keV due to a $^{54}$Mn $\gamma$ line is also identified:
the presence of $^{54}$Mn stems from the cosmogenic activation of copper~\cite{CUORE:2016ons,Laub2009}.
The visible peak to the right of Q$_{\beta\beta}$ is the 911.2\,keV $\gamma$ line from $^{228}$Ac,
another element of the $^{232}$Th chain.
In addition, we observe a continuous background contribution mainly induced by the 2$\nu\beta\beta$ decay of $^{130}$Te
and by multiple Compton scattering of the various $\gamma$ rays from environmental radioactivity and cosmic radiation.
The choice of the ROI is driven by the need for the energy window to fully contain the events of the posited \nbb peak,
while being large enough to include and constrain the background structures, allowing us to evaluate the signal rate correctly.
The ROI contains the $^{54}$Mn and $^{208}$Tl peaks, while the $^{228}$Ac line is excluded as it is 45\,keV
($>5\sigma$ with FWHM energy resolution of $\sim 4.3$\,keV in the ROI) away from Q$_{\beta\beta}$.

We perform a simultaneous binned Bayesian fit on the five included datasets.
The fit is performed with the Bayesian Analysis Toolkit (BAT)~\cite{Caldwell:2008fw},
that samples from the posterior probability density by performing a Markov Chain Monte Carlo (MCMC)
using the Metropolis-Hastings algorithm. We fit the CUORE ${\cal{M}}_1$ spectrum over the chosen ROI;
the lower limit on the 0$\nu\beta\beta$ decay rate is taken as the rate corresponding to 90\% of the marginalized posterior.

We fit the CUORE ${\cal{M}}_1$ spectrum over the chosen ROI with a likelihood that 
includes the posited signal peak plus the background structures present in the ROI,
namely the $^{54}$Mn peak, the $^{208}$Tl peak, and a continuum distribution.
We model the latter with a linear function, that describes the decreasing trend over the fit region. This simpler effective model is consistent with the full CUORE background model~\cite{CUORE:2020bok}.
The binned likelihood for each dataset is the product of Poisson terms, and the total likelihood is:
\begin{equation}
  \mathscr{L} =\prod_{ds} \prod_{i}^{N_{bins}} \frac{\mu_i^{n_i}\,e^{-\mu_i}}{n_i!}\quad,
  \label{eq:binned_likelihood}
\end{equation}
where $ds$ indexes the dataset, and the index $i$ runs over the 140 bins (0.5\,keV/bin).
In the approximation of small bin width, the number of expected counts $\mu_i$ in the $i$-th bin can be taken as the value of the model function at the center of the bin: 
\begin{equation}
  \mu_i = Sf_S^{ds}(i) + C_{\text{Mn}}f_{\text{Mn}}^{ds}(i) +
  C_{\text{Tl}}f_{\text{Tl}}^{ds}(i) + f^{ds}_{\text{linear}}(i)\ \ ,
  \label{eq:mu_i_components}
\end{equation}
where $S$, $C_{\text{Mn}}$ and $C_{\text{Tl}}$ are the number of counts at the signal, $^{54}$Mn and $^{208}$Tl peaks,
while $f_S^{ds}(i)$, $f_{\text{Mn}}^{ds}(i)$, $f_{\text{Tl}}^{ds}(i)$, 
and $f^{ds}_{\text{linear}}(i)$ are the values at the $i$-th bin of the probability density functions used to model the shape of each component.

We model the shape of each peak as the sum of three Gaussian distributions based on the CUORE detector response function, corrected for the energy dependence of the detector response (energy-resolution scaling and energy reconstruction bias) studied in Ref~\cite{CUORE:2019yfd}.
The definition of each component of Eq.~\ref{eq:mu_i_components} is detailed in the following.
We implement all terms as parameters of the fit.

The 0$\nu\beta\beta$ decay rate $\Gamma_{0\nu}$ is connected to the expected number of signal counts $S$
for a given dataset through the formula:
\begin{equation}
  S = \Gamma_{0\nu}\cdot \frac{N_A}{A_{\text{TeO}_2}}\cdot \eta_{128} \cdot (M\Delta t)_{ds} \cdot \epsilon^{cut}_{ds} \cdot \epsilon_{\text{MC}}\quad,
\end{equation}
where $N_A$ is Avogadro's constant, $A_{\text{TeO}_2}$ is the TeO$_2$ molar mass, $\eta_{128}$ is the $^{128}$Te natural isotopic abundance, $(M\Delta t)_{ds}$ is the dataset exposure (in units of\,kg$\cdot$yr), $\epsilon^{cut}_{ds}$ is the dataset total analysis efficiency, and $\epsilon_{\text{MC}}$ is the containment efficiency. The decay rate $\Gamma_{0\nu}$ in the model is a global parameter common to all the datasets.
We make a statistical inference on this parameter of interest.

The $^{54}$Mn originates from cosmogenic activation of Cu,
which occurred before the CUORE cryostat and detector structure components were moved underground at LNGS.
This element has a half-life of 312.2 days; the analyzed data were taken over a period of $\sim$2 years,
thus we expect the number of events due to $^{54}$Mn decay  to decrease over time.
To account for this reduction, we include a multiplicative factor
in the definition of the number of expected $^{54}$Mn events in each dataset, resulting from the integration of the exponential decay over the dataset duration:
\begin{equation}
  C_{\text{Mn}} = \Gamma_{\text{Mn}}
  \cdot \tau \cdot (e^{-\frac{t^{\mathrm{in}}_{ds}}{\tau_{\text{Mn}}}} - e^{-\frac{t^{\mathrm{fin}}_{ds}}{\tau_{\text{Mn}}}} ) \cdot \frac{(M\Delta t)_{ds}}{t^{\mathrm{fin}}_{ds} - t^{\mathrm{in}}_{ds}} \cdot \epsilon^{cut}_{ds}\quad,
  \label{eq:MnRate_likel}
\end{equation}
where $t^{\mathrm{in}}_{ds}$ and $t^{\mathrm{fin}}_{ds}$ respectively refer to the start-time and the end-time of the dataset with respect to the beginning of the data taking. The livetime fraction $\frac{\Delta t_{ds}}{t^{\mathrm{fin}}_{ds} - t^{\mathrm{in}}_{ds}}$ accounts for the dead times - few time intervals of data taking that are removed from the analysis, for example noisy periods due to short maintenance interruptions, activities in the local laboratory or earthquakes - over the integration time interval. 
The $^{54}$Mn rate $\Gamma_{\text{Mn}}$ (units of counts/(kg$\cdot$yr))
is a nuisance parameter of the fit common to all the datasets.

$^{208}$Tl belongs to the naturally occurring $^{232}$Th chain.
Given that the amplitude of the observed higher intensity $^{208}$Tl $\gamma$ peaks
are constant in time across the datasets, we assume the 860.6\,keV rate to be stable.
We then define the expected number of events at this $^{208}$Tl line in the ROI for a given dataset as:
\begin{equation}
  C_{\text{Tl}} = \Gamma_{\text{Tl}}\cdot (M\Delta t)_{ds} \cdot \epsilon^{cut}_{ds}\quad,
  \label{eq:TlRate_likel}
\end{equation}
where the $^{208}$Tl decay rate $\Gamma_{\text{Tl}}$ is expressed in units of counts/(kg$\cdot$yr).
As with the $^{54}$Mn rate, this represents a nuisance parameter of the fit.

We model the continuous background distribution as a linear function of energy according to the following expression:
\begin{equation}
  f^{ds}_{\text{linear}}(i) = C^{ds}_b + m_{ds} (E_i - E_{1/2})\quad,
  \label{eq:128bkg_term}
\end{equation}
where $C^{ds}_b$ and $m_{ds}$ are the expected number of background events
and the background slope for a given dataset, $E_i$ is the energy at the center of the $i$-th bin,
and $E_{1/2}$ is the energy corresponding to center of the ROI.
We define the expected number of events $C^{ds}_b$ in each dataset as:
\begin{equation}
  C^{ds}_b = \text{BI}_{ds} \cdot (M\Delta t)_{ds} \cdot w_i\quad,
  \label{eq:BI_likel}
\end{equation}
where BI$_{ds}$ is the background index of dataset $ds$ in units of counts/(keV$\cdot$kg$\cdot$yr)
and $w_i$ is the bin width, which is constant across the energy spectrum.
The slope and the background index are also nuisance parameters in the fit and are dataset-dependent quantities.

We adopt a uniform prior for each parameter of the fit for several reasons.
Due to the 100-fold increase in exposure, CUORE's sensitivity on $\Gamma_{0\nu}$ is expected to be factor of $\sim$10 better with respect to the past direct limit. The absence of knowledge on $\Gamma_{0\nu}$ at the range that CUORE can probe justifies the choice of a uniform prior for $\Gamma_{0\nu}\ge 0$ according to the Principle of Indifference, which assigns equal probabilities to all the possible values up to a maximum that can be greater that the past limit. The CUORE Background Model can provide information on some nuisance parameters, however it is constructed through a fit on the same data
that are used for the present analysis,
and including such information would bias the result.
Thus, in the absence the of independent measurements, we use
a uniform prior for all nuisance parameters.
The signal, $^{54}$Mn and $^{208}$Tl rates and the BI are constrained to non-negative physical values only,
while for the background slope $m$ both negative and positive values are allowed.
\begin{figure}[h]
  \centering
  \includegraphics[width=1.\columnwidth]{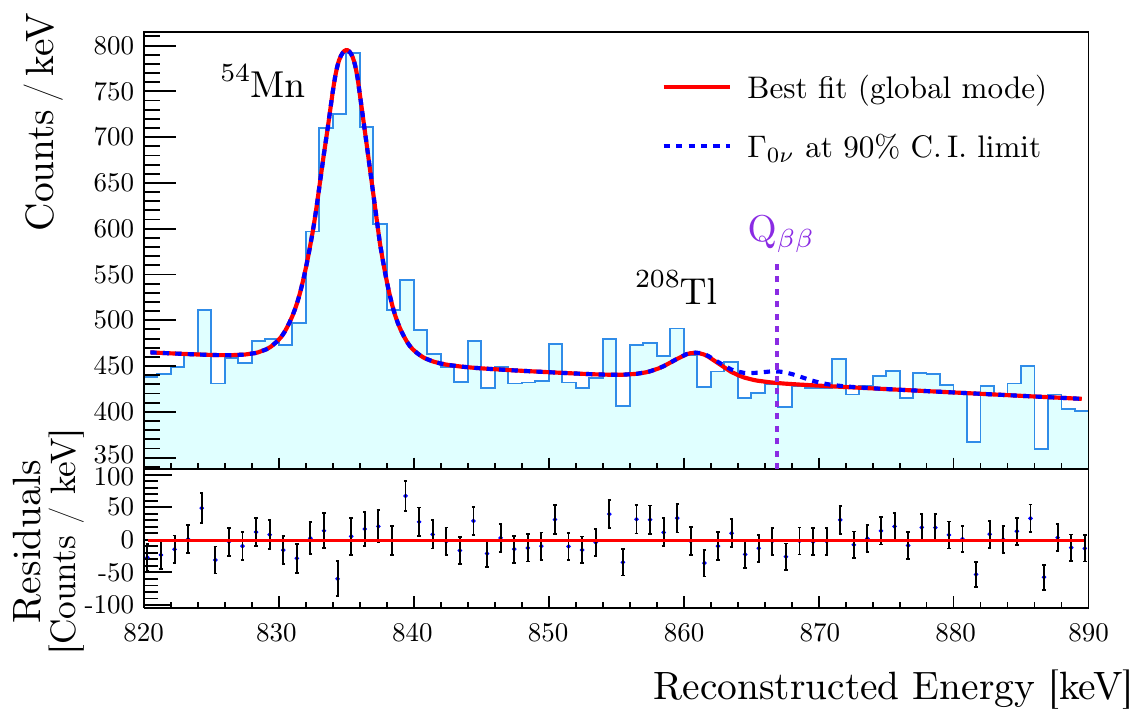}
  \caption{Top: data spectrum in the ROI, together with the best-fit curve (red solid) and the best-fit curve with the signal rate set at the 90\% CI limit (blue dashed). Bottom: residual plot with fit, compatible with 0 (intercept at $-1.1\pm2.6$ counts/keV, $\chi^2/\mathrm{dof}=82/69$.)}
  \label{fig:SBpos_fit}
\end{figure}
We run the Bayesian fit on the data, and find no evidence for $^{128}$Te 0$\nu\beta\beta$ decay.
From the marginalized posterior distribution of the signal rate,
we extract a 90\% CI limit of
\begin{equation}
  \Gamma_{0\nu} < 1.9\cdot 10^{-25}\,\, \mathrm{yr}^{-1}\quad.
\end{equation}
This lower limit corresponds to a 90\% CI upper limit on the $^{128}$Te 0$\nu\beta\beta$ decay half-life of
\begin{equation}
  \mathrm{T}^{0\nu}_{1/2} > 3.6\cdot 10^{24}\,\, \mathrm{yr}\quad.
\end{equation}
This result is the most stringent limit on the \nbb decay of $^{128}$Te to date, representing a more than 30-fold improvement over the previous limit~\cite{Arnaboldi:2002te} from direct searches,
and exceeds for the first time the combined \nbb and \vvbb decay half life obtained by geochemical measurements. The fit result and the total ROI spectrum are shown in Fig.~\ref{fig:SBpos_fit}.

\begin{figure}[h]
  \centering
  \includegraphics[width=1.\columnwidth]{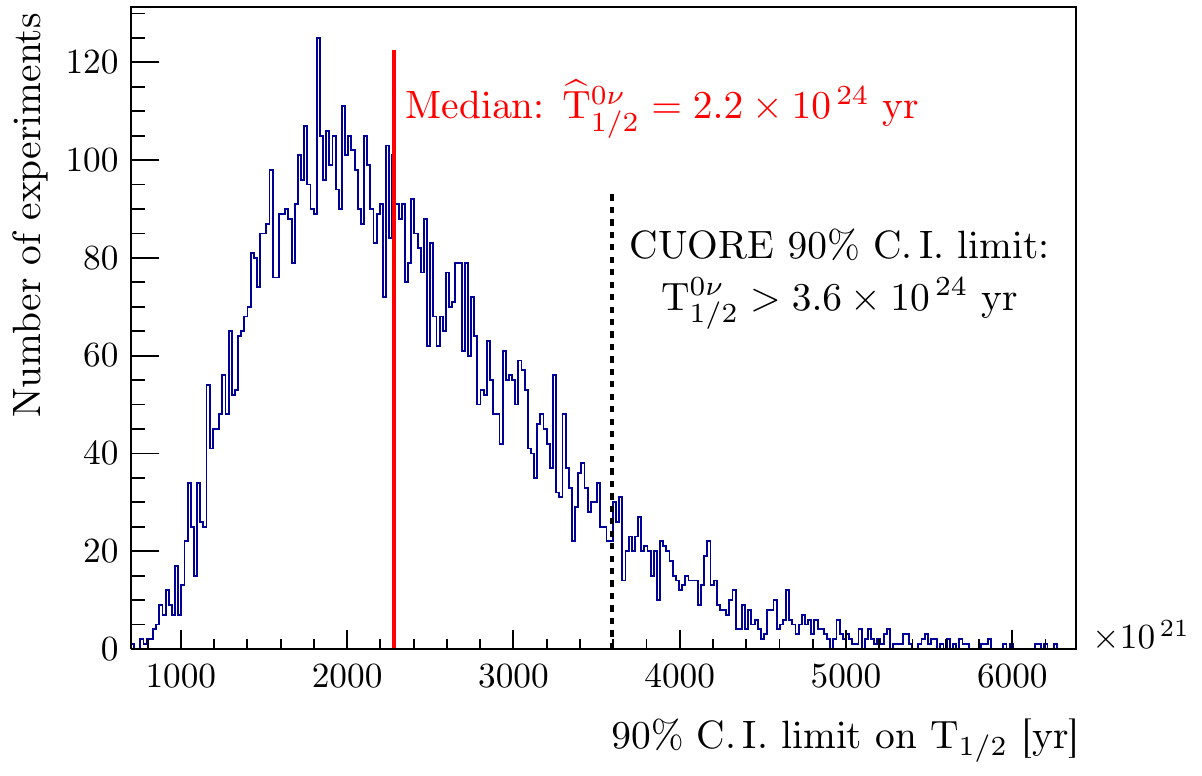}
  \caption{Distribution of the 90\% CI limits on T$^{0\nu}_{1/2}$ extracted from repeating the analysis on the $10^4$ background-only pseudo-experiments. 
    The solid line corresponds to the median exclusion sensitivity, while the dashed one shows the 90\% CI limit from the analysis of the CUORE data.
    }
  \label{fig:sensitivity}
\end{figure}

We extract the median exclusion sensitivity to $^{128}$Te \nbb decay
by repeating the statistical only Bayesian fit on 10$^4$ toy-MC simulations of the 
experiment.
We produce the toy-MCs using the global mode values of the background parameters
from a Bayesian fit without the signal component on the CUORE data.
The median exclusion sensitivity is the median of the distribution of the 90\% CI limits on T$^{0\nu}_{1/2}$,
each resulting from a signal plus background fit to one of the 10$^4$ background-only toy-MCs.
This distribution is shown in Fig.~\ref{fig:sensitivity}, and its median is
$\hat{\mathrm{T}}^{0\nu}_{1/2} = 2.2\cdot 10^{24}\,\, \mathrm{yr}$.
The probability to obtain a more stringent limit than the one observed with the CUORE data is 8.8\%.
We also repeat the fit on the data, 
allowing the signal rate to assume non-physical negative values.
In this case, the global mode of $\Gamma_{0\nu}$ is \mbox{$(-2.4\pm1.8)\cdot 10^{-25}$ yr$^{-1}$,}
resulting in an under-fluctuation with a statistical significance of $\sim 1.4 \sigma$,
which is compatible with the 8.8\% under-fluctuation obtained from the sensitivity study.

We summarize in Table~\ref{tab:systematics} a series of systematic uncertainties
affecting our limit.
For this study, we run the fit without the constraint $\Gamma_{0\nu}\ge 0$,
to access the full range $\Gamma_{0\nu}$ marginalized posterior.
We adopt a fully Bayesian approach to evaluate the effect
due to the uncertainties on the containment efficiency,
the analysis cut efficiency and the $^{128}$Te natural isotopic abundance.
We implement these as independent nuisance parameters in the likelihood with a Gaussian prior,
whose mean and sigma are equal to the respective central value and associated error.
We thus repeat the Bayesian fit activating one nuisance parameter
at the time to allow its value to vary according to the corresponding prior. 

We treat the systematics due to the uncertainty on the $^{128}$Te Q$_{\beta\beta}$
and on the detector response function parameters (namely the energy reconstruction bias and resolution scaling)
using an alternative approach, which we refer to as the Repeated Fit Approach,
because of the excessive computation time required 
to treat them as nuisance parameters in the fit.
This method consists of repeating the fit for a series of discrete values
of the systematic parameter under study, covering a $\pm3\sigma$ region around its prior mean value.
We then sum the $\Gamma_{0\nu}$ marginalized posteriors obtained
from each fit weighting by the prior probability of the parameter considered as systematic,
and take the signal rate corresponding to the 90\% quantile of the obtained distribution.
We take additional care when treating the detector response parameter systematics.
It was previously observed in CUORE~\cite{CUORE:2019yfd} that both the bias
on the energy reconstruction and the resolution scaling exhibit an energy dependence
which we model with two independent second order polynomial functions.
As a consequence, a set of three correlated parameters describes
the energy bias and another set of three exists for the resolution scaling.
The correlations among these parameters are taken into account
using multi-dimensional priors.
The dominant systematic is Q$_{\beta\beta}$, which has an effect of 7.0\% on the limit.
We expect this due to the relatively large error on its literature value, ($866.7\pm0.7)$\,keV~\cite{Wang:2021xhn}.
All the other systematics affect the limit by less than 1\%;
the $^{128}$Te isotopic abundance, the analysis cut efficiency,
and the detector response function parameters result in values below
the intrinsic BAT uncertainty due to the MCMC stochastic behavior (0.3\%).

\begin{table}
  \caption{Systematic effects on the $^{128}$Te \nbb decay signal rate 90\% limit.
    The first row refers to the BAT intrinsic uncertainty, due to the stochastic nature of the MCMC.
    The efficiencies and the $^{128}$Te isotopic abundance were treated as nuisance parameters in the fit,
    while an alternative approach was adopted for the uncertainties on the $^{128}$Te Q$_{\beta\beta}$ and the detector response function parameters.
    The dominant systematic effect is the value on Q$_{\beta\beta}$.}
  \label{tab:systematics}
  \centering
  \begin{tabular}{c@{\hspace{4mm}}c@{\hspace{4mm}}c}
    \hline
    \hline
    Systematic & Prior & Effect on $\Gamma_{0\nu}^{90\%}$\\
    \hline
    BAT Stat. Only fit & - & 0.3\% \\
    \hline
    \multicolumn{3}{c}{Bayesian Approach} \\
    \hline
    Containment Efficiency & Gaussian & 0.4\% \\
    $^{128}$Te Isotopic Abundance & Gaussian & $<$0.3\% (0.05\%) \\ 
    Analysis Cut Efficiency & Gaussian & $<$0.3\% (0.1\%) \\
    \hline
    \multicolumn{3}{c}{Repeated Fit Approach} \\
    \hline
    $^{128}$Te Q$_{\beta\beta}$ & Gaussian & 7.0\% \\
    Energy Reconstruction Bias & Multivariate & $<$0.3\% (0.1\%) \\
    Energy Resolution Scaling & Multivariate & $<$0.3\% (0.1\%) \\
    
    \hline
    \hline
  \end{tabular}
\end{table}

Several Standard Model extended theories include mechanisms that try to explain how the \nbb decay takes place~\cite{Pontecorvo:1968wp,Mitra:2011qr,Tello:2010am}.
Among these models, the exchange of a light Majorana neutrino is the most favored~\cite{Dell'Oro:2016dbc}. However, a positive signal for the \nbb decay of a single isotope would not determine the mechanism of this process~\cite{Bilenky:2014uka}. Discriminating among the existing models~\cite{Deppisch:2006hb} and possibly testing the calculations of nuclear matrix elements for double beta decay~\cite{Bilenky:2002ga}, would be possible via the comparison of results from different isotopes. It has been pointed out~\cite{Deppisch:2006hb} that the study of the \nbb decay of $^{128}$Te can be particularly useful for such model discrimination.
In this paper, we present the first results on the $^{128}$Te \nbb decay search with the CUORE experiment.
With a binned Bayesian fit of the CUORE data with a total exposure of 309.33\,kg$\cdot$yr (78.6\,kg$\cdot$yr of $^{128}$Te),
we find no evidence for $^{128}$Te \nbb decay, and we set a 90\% CI limit on the half-life
of this process at T$^{0\nu}_{1/2}>3.6\cdot 10^{24}$ yr.
This represents the most stringent limit in literature,
improving by over a factor 30 the previous limit from a direct search experiment,
and exceeding those from indirect geochemical measurements for the first time.
From the analyzed exposure, the CUORE median exclusion sensitivity to this decay
is $\hat{\mathrm{T}}^{0\nu}_{1/2} = 2.2\cdot 10^{24}$ yr,
giving an 8.8\% probability to obtain a stronger limit.
The dominant systematic, affecting the result at the level of 7.0\%, is due to the uncertainty on Q$_{\beta\beta}$. 

The analysis presented in this paper has been carried out with about one-tenth of the final exposure scheduled for CUORE, corresponding to $\sim$3.7 ton$\cdot$y. We plan to update these results with such unprecedented amount of data collected with TeO$_2$ crystals.

\begin{acknowledgments}
  The CUORE Collaboration thanks the directors and staff of the Laboratori Nazionali del Gran Sasso and the technical staff of our laboratories. This work was supported by the Istituto Nazionale di Fisica Nucleare (INFN); the National Science Foundation under Grant Nos. NSF-PHY-0605119, NSF-PHY- 0500337, NSF-PHY-0855314, NSF-PHY-0902171, NSF- PHY-0969852, NSF-PHY-1307204, NSF-PHY-1314881, NSF-PHY-1401832, and NSF-PHY-1913374; and Yale University. This material is also based upon work supported by the US Department of Energy (DOE) Office of Science under Contract Nos. DE-AC02- 05CH11231 and DE-AC52-07NA27344; by the DOE Office of Science, Office of Nuclear Physics under Contract Nos. DE-FG02-08ER41551, DE-FG03-00ER41138, DE- SC0012654, DE-SC0020423, DE-SC0019316; and by the EU Horizon2020 research and innovation program under the Marie Sklodowska-Curie Grant Agreement No. 754496. This research used resources of the National Energy Research Scientific Computing Center (NERSC). This work makes use of both the DIANA data analysis and APOLLO data acquisition software packages, which were developed by the CUORICINO, CUORE, LUCIFER and CUPID-0 Collaborations.

\end{acknowledgments}

\bibliography{ref}

\section{Supplemental Material}

We develop and optimize the fit strategy on toy-MC spectra.
We generate the toy-MCs according to the signal-plus-background model,
extracting the values of $\Gamma_{\text{Mn}}$, $\Gamma_{\text{Tl}}$, BI and $m$
from the CUORE Background Model.
We refer to Ref. \cite{Dompe:PhD2021} for a more detailed discussion of the method.
We take advantage of the toy-MCs to verify that the fit correctly reconstructs
the simulated background components and to inspect if a bias is introduced
in the \nbb decay rate reconstruction when a signal contribution is added in the toy-MC. 
We generate $10^4$ toy-MCs with no signal and run the fit independently on each of them.
We then construct the distributions of the best-fit values from all the toy-MCs for each parameter,
in order to compare the extracted and simulated values.
As expected, these distributions are centered at the values used to produce the toy-MC.
Thanks to the large number of toy-MCs, we are able to identify a small bias
in the reconstruction of the BI and the slope corresponding
to a $<$0.15\% underestimation and a $\le$1.6\% overestimation, respectively.
No correlations are seen between these two parameters.
The reconstructed values of the $^{54}$Mn and $^{208}$Tl rates are compatible with the injected values.
To test the signal rate reconstruction, we repeat the fit on five sets of 2000 toy-MCs,
injecting a different signal amplitude in the range (2~--~10)$ \cdot 10^{-25}$ yr$^{-1}$ in each set. 
This range includes the signal rate corresponding to the CUORE sensitivity of
$3.2\cdot 10^{-25}$ yr$^{-1}$ obtained from pure toy-MC, i.\,e.\ without including real data.

\begin{figure}[htbp]
  \centering
  \includegraphics[width=1.\columnwidth]{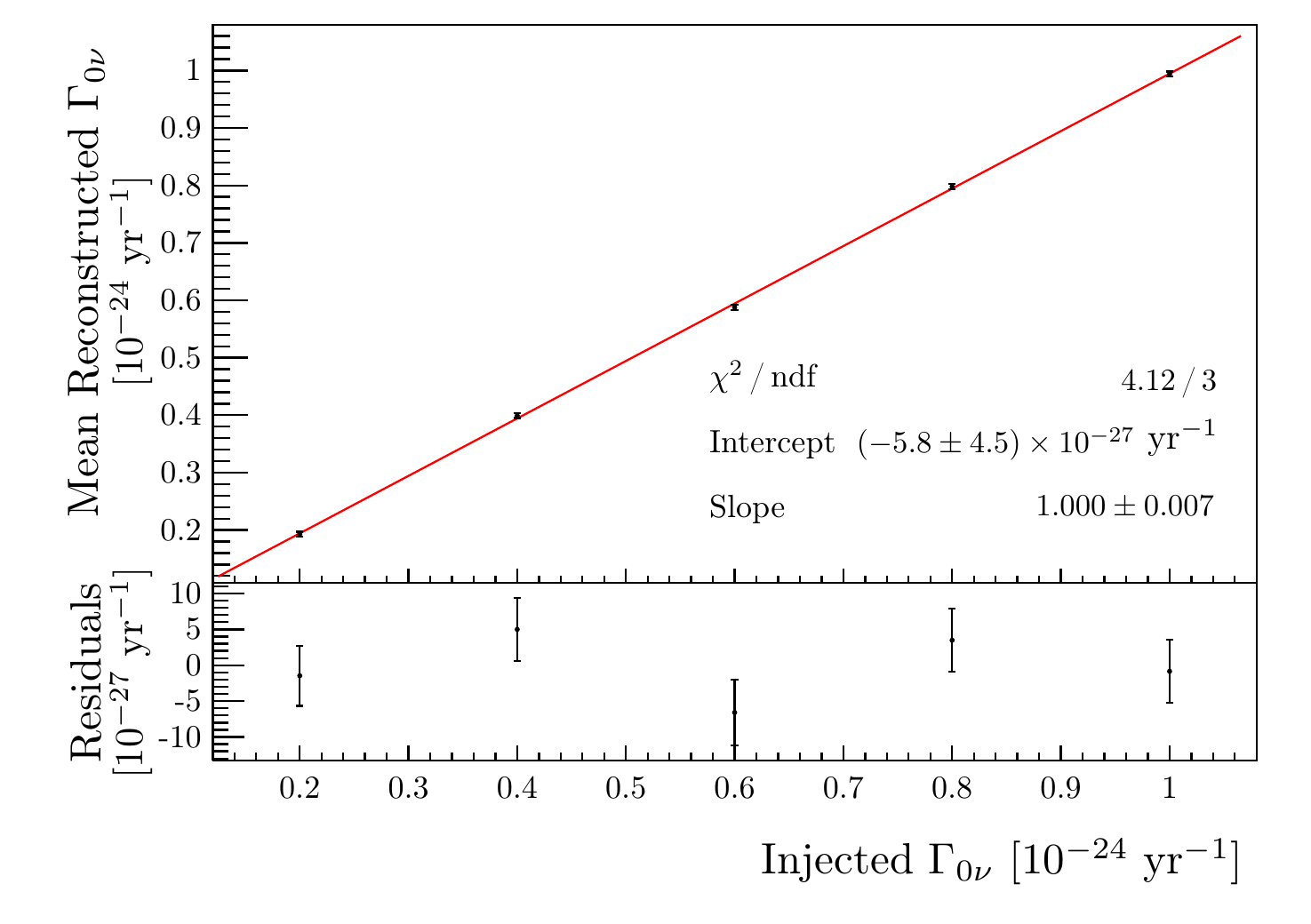}
  \caption{Linear fit on the mean reconstructed signal rate as a function of the injected one.
    The intercept and slope are compatible with 0 and 1, respectively.}
  \label{fig:recoVSinj_rate}
\end{figure}

Figure~\ref{fig:recoVSinj_rate} shows the mean reconstructed signal rate as a function of the injected one.
The relation between the two is well described by a linear function:
the intercept is compatible with 0 at a $\sim1.3\,\sigma$,
and the slope is compatible with 1 within $1\sigma$.
These results allow us to conclude that no bias is introduced
by the Bayesian fit in the signal rate reconstruction.

We also study the intrinsic stability of the BAT fit,
by repeating it 2$\cdot10^3$ times on the same toy-MC populated only with the background components,
obtaining a $0.3$\% root mean square on the distribution of the $\Gamma_{0\nu}$ limits at 90\% credibility interval (CI).

Table~\ref{tab:prior_ranges} reports the ranges for all fit parameters.
All parameters proportional to a number of counts, namely the signal, Mn and Tl rates, are allowed to assume only non-negative values.
The BI of each dataset is further constrained according to a preliminary estimation of the number of background counts.
The background slopes are allowed to assume also negative values.

Table~\ref{tab:results_s+b_positiveRate} reports the value at the global mode for all parameters
of the fit to the data.

\begin{figure}[H]
  \centering
  \includegraphics[width=1.\columnwidth]{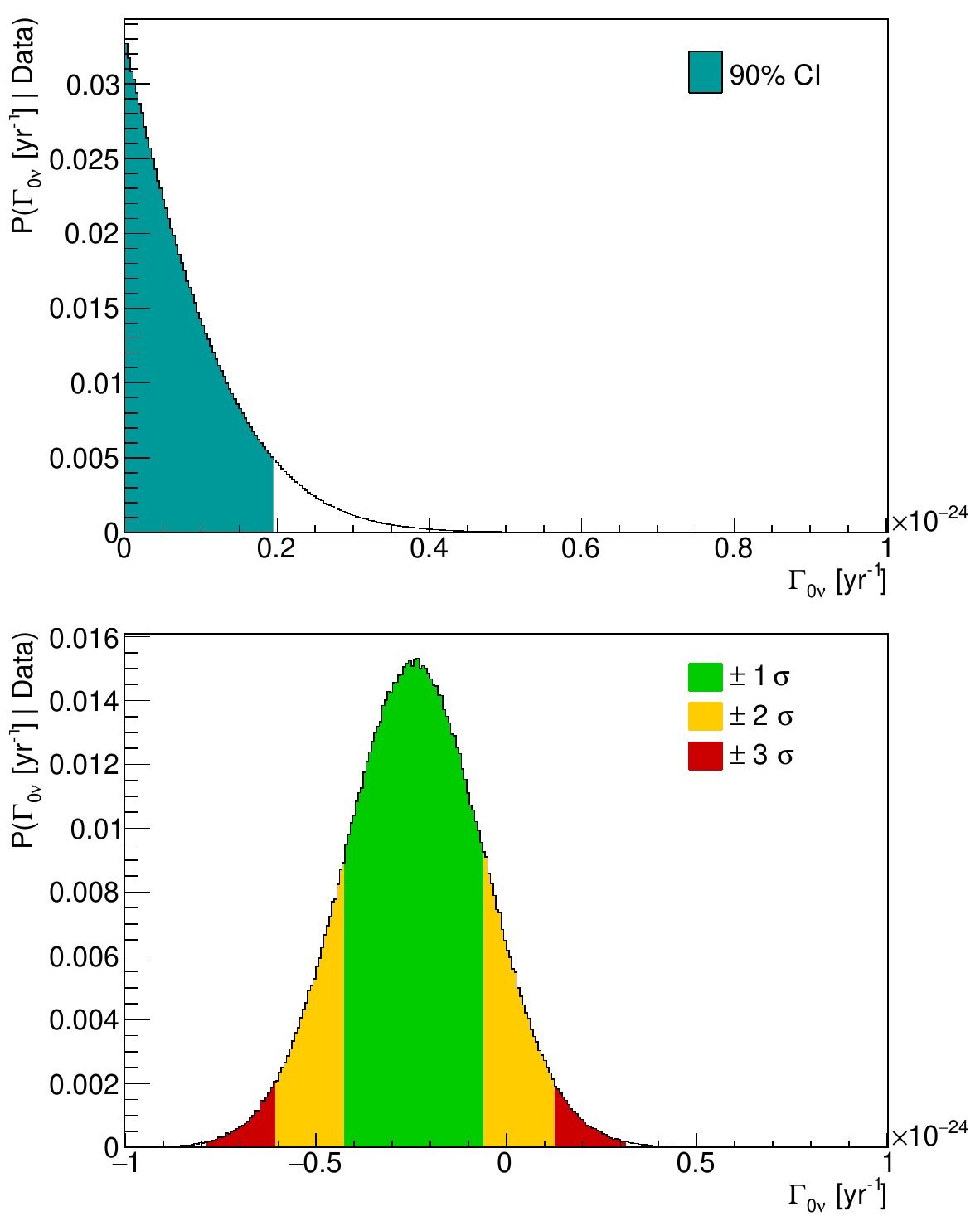}
  \caption{Top: marginalized posterior of the signal rate obtained from the official fit, which allows only physical values.
    Bottom: marginalized posterior of the signal rate obtained from the alternative fit, which allows also non-physical values of $\Gamma_{0\nu}$.
    A $\sim 1.4\sigma$ significance under-fluctuation, compatible with the results of the sensitivity studies, is observed.
    Both distributions are normalized to one.}
  \label{fig:rate_posterior}
\end{figure}

Figure~\ref{fig:rate_posterior} shows the posterior distribution
for $\Gamma_{0\nu}$ obtained from the reference fit, and from the alternative fit
performed with the signal rate allowed to artificially assume non-physical negative values.

\begin{table}[h]
  \caption{Parameter ranges for the fit parameters.
    All parameters are assigned a uniform prior;
    the signal, Mn and Tl rates and the BI are constrained to non-negative physical values only,
    while both positive and negative values are allowed for the background slope.}
  \label{tab:prior_ranges}
  \centering
  \begin{tabular}{c@{\hspace{8mm}}c@{\hspace{8mm}}c}
    \hline
    \hline
    \multirow{1}{*}{Parameter} & Prior Range \\
    \hline
    $\Gamma_{0\nu}$
    & [0, 1.74$\cdot 10^{-24}$] yr$^{-1}$ \\
    
    $BI_1$ & [1.1634, 1.73] cts/(keV$\cdot$kg$\cdot$yr)\\
    
    $BI_2$ & [1.188, 1.6513] cts/(keV$\cdot$kg$\cdot$yr)\\
    
    $BI_3$ & [1.2453, 1.7374] cts/(keV$\cdot$kg$\cdot$yr)\\
    
    $BI_4$ & [1.2204, 1.7412] cts/(keV$\cdot$kg$\cdot$yr)\\
    
    $BI_5$ & [1.0221, 1.4536] cts/(keV$\cdot$kg$\cdot$yr)\\
    
    $m_1$ & [-1, 1] 1/keV\\
    
    $m_2$ & [-1, 1] 1/keV\\
    
    $m_3$ & [-1, 1] 1/keV\\
    
    $m_4$ & [-1, 1] 1/keV\\
    
    $m_5$ & [-1, 1] 1/keV\\
    
    $\Gamma_{\text{Mn}}$ & [0, 44.58] cts/(kg$\cdot$yr)\\
    
    $\Gamma_{\text{Tl}}$ & [0, 6.16] cts/(kg$\cdot$yr)\\
    \hline
    \hline
  \end{tabular}
\end{table}

\newpage

\begin{table}[h]
  \caption{Best-fit values for all parameters of the fit on CUORE data.
    The signal rate is allowed to take non-negative values only.}
  \label{tab:results_s+b_positiveRate}
  \centering
  \begin{tabular}{c@{\hspace{8mm}}c@{\hspace{8mm}}c}
    \hline
    \hline
    \multirow{1}{*}{Parameter} & Fit Result & Units \\
    \hline
    $\Gamma_{0\nu}$ & 
    0 & yr$^{-1}$\\
    
    $BI_1$ & $1.48 \pm 0.02$ & cts/(keV$\cdot$kg$\cdot$yr)\\
    
    $BI_2$ & $1.43 \pm 0.02$ & cts/(keV$\cdot$kg$\cdot$yr)\\
    
    $BI_3$ & $1.49\pm 0.02$ & cts/(keV$\cdot$kg$\cdot$yr)\\
    
    $BI_4$ & $1.48\pm 0.02$ & cts/(keV$\cdot$kg$\cdot$yr)\\
    
    $BI_5$ & $1.26 \pm 0.02$ & cts/(keV$\cdot$kg$\cdot$yr)\\
    
    $m_1$ & $-0.07\pm 0.03$ & keV$^{-1}$\\
    
    $m_2$ & $-0.06\pm 0.03$ & keV$^{-1}$\\
    
    $m_3$ & $-0.08\pm 0.03$ & keV$^{-1}$\\
    
    $m_4$ & $-0.04\pm 0.03$ & keV$^{-1}$\\
    
    $m_5$ & $-0.12\pm 0.03$ & keV$^{-1}$\\
    
    $\Gamma_{\text{Mn}}$ & $15.9\pm 0.7$ & cts/(kg$\cdot$yr)\\
    
    $\Gamma_{\text{Tl}}$ & $0.5\pm 0.2$ & cts/(kg$\cdot$yr)\\
    \hline
    \hline
  \end{tabular}
\end{table}

\end{document}